\documentclass{aastex631}

\begin{document}

\title{\textit{At First Sight!} Zero-Shot Classification of Astronomical Images with Large Multimodal Models}

\correspondingauthor{Dimitrios Tanoglidis}
\email{tanoglidisdimitrios@gmail.com}

\author[0000-0002-4631-4529]{Dimitrios Tanoglidis}
%Washington, DC 20006, USA}

\author[0000-0002-8220-3973]{Bhuvnesh Jain}
\affiliation{Department of Physics and Astronomy, University of Pennsylvania, Philadelphia, PA 19104, USA}
%\affiliation{}

%% Note that the \and command from previous versions of AASTeX is now
%% depreciated in this version as it is no longer necessary. AASTeX 
%% automatically takes care of all commas and "and"s between authors names.

%% AASTeX 6.31 has the new \collaboration and \nocollaboration commands to
%% provide the collaboration status of a group of authors. These commands 
%% can be used either before or after the list of corresponding authors. The
%% argument for \collaboration is the collaboration identifier. Authors are
%% encouraged to surround collaboration identifiers with ()s. The 
%% \nocollaboration command takes no argument and exists to indicate that
%% the nearby authors are not part of surrounding collaborations.

%% Mark off the abstract in the ``abstract'' environment. 
\begin{abstract}

Vision-Language multimodal Models (VLMs) offer the possibility for zero-shot classification in astronomy: i.e. classification via natural language prompts, with no training. 
We investigate two models, GPT-4o and LLaVA-NeXT, for zero-shot classification of low-surface brightness galaxies and artifacts, as well as morphological classification of galaxies. We show that with natural language prompts these models achieved significant accuracy (above 80 percent typically) without additional training/fine tuning. 
We discuss areas that require improvement, especially for LLaVA-NeXT, which is an open source model. 
Our findings aim to motivate the astronomical community to consider VLMs as a powerful tool for both research and pedagogy, with the prospect that future custom-built or fine-tuned models could perform  better.

\end{abstract}

%% Keywords should appear after the \end{abstract} command. 
%% The AAS Journals now uses Unified Astronomy Thesaurus concepts:
%% https://astrothesaurus.org
%% You will be asked to selected these concepts during the submission process
%% but this old "keyword" functionality is maintained in case authors want
%% to include these concepts in their preprints.

%% From the front matter, we move on to the body of the paper.
%% Sections are demarcated by \section and \subsection, respectively.
%% Observe the use of the LaTeX \label
%% command after the \subsection to give a symbolic KEY to the
%% subsection for cross-referencing in a \ref command.
%% You can use LaTeX's \ref and \label commands to keep track of
%% cross-references to sections, equations, tables, and figures.
%% That way, if you change the order of any elements, LaTeX will
%% automatically renumber them.
%%
%% We recommend that authors also use the natbib \citep
%% and \citet commands to identify citations.  The citations are
%% tied to the reference list via symbolic KEYs. The KEY corresponds
%% to the KEY in the \bibitem in the reference list below. 

\section{Introduction} \label{sec:intro}

Multi-modal foundation models have the potential to impact diverse scientific fields. In astronomy, the vast and varied forms of data offer opportunities. From images taken at different wavelengths; spectra and time domain observations; and in some limited cases language labels. Given these potential opportunities, there have been a few attempts at multi-modal investigations in astronomy \citep[e.g.,][]{astroclip,Sharma_2024}

%The data can be images taken in different optical filters, as well as images at shorter and longer wavelengths with fundamentally different telescopes; spectra with high resolution spectrographs, as well as spectra with grisms and other lower resolution instruments that are qualitatively different; time domain observations; and in some limited cases language labels. The language labels can take diverse forms, such as identified spectra lines and human classification labels of galaxy types. Given these potential opportunities, there have been a few attempts at multi-modal investigations but by no means the kind of ambitious models that have been applied to climate modeling. 
%BJ: can add references from our transformer write up. also need to think a bit more about the long para about images spectra and language, i'm probably missing some things
%LN: In astronomy, the vast and varied forms of data offer opportunities to utilize these models. (something like this at the end to make the sentence feel more complete?)

Despite the recent impact of deep learning in astronomy, especially in image classification tasks, a significant challenge remains: the lack of large, high-quality, labeled datasets necessary for the training of supervised deep learning models. For each specific task (e.g., morphological classification of galaxies) a separate model has to be trained using labeled data for that task. Furthermore, data obtained through one instrument or telescope cannot be readily used to train a model to perform the \textit{same} task using data from another instrument. In such cases, domain adaptation techniques should be employed \citep[e.g.,][]{Ciprijanovic_2022}.

Foundation models \citep{Foundation_Models}, on the other hand, are models trained on broad datasets,  that can subsequently be adapted in a range of downstream tasks, such as classification. The case where the model can categorize objects or classes that have never been seen before, is referred to as \textit{zero-shot} learning. In this research note, we explore the potential of large, multimodal, Vision-Language models (VLMs; see for example \citet{Vision_Language}) to perform zero-shot classification of astronomical images, by \textit{leveraging natural language descriptions of the different object classes}. Our results, using generalist models that have not been fine-tuned on astronomical data, serve as a proof of concept for the potential of VLMs in astronomy. 

\section{METHODS} \label{sec:methods}

\begin{figure}[!ht]
    \centering
    \includegraphics[width=6.8cm]{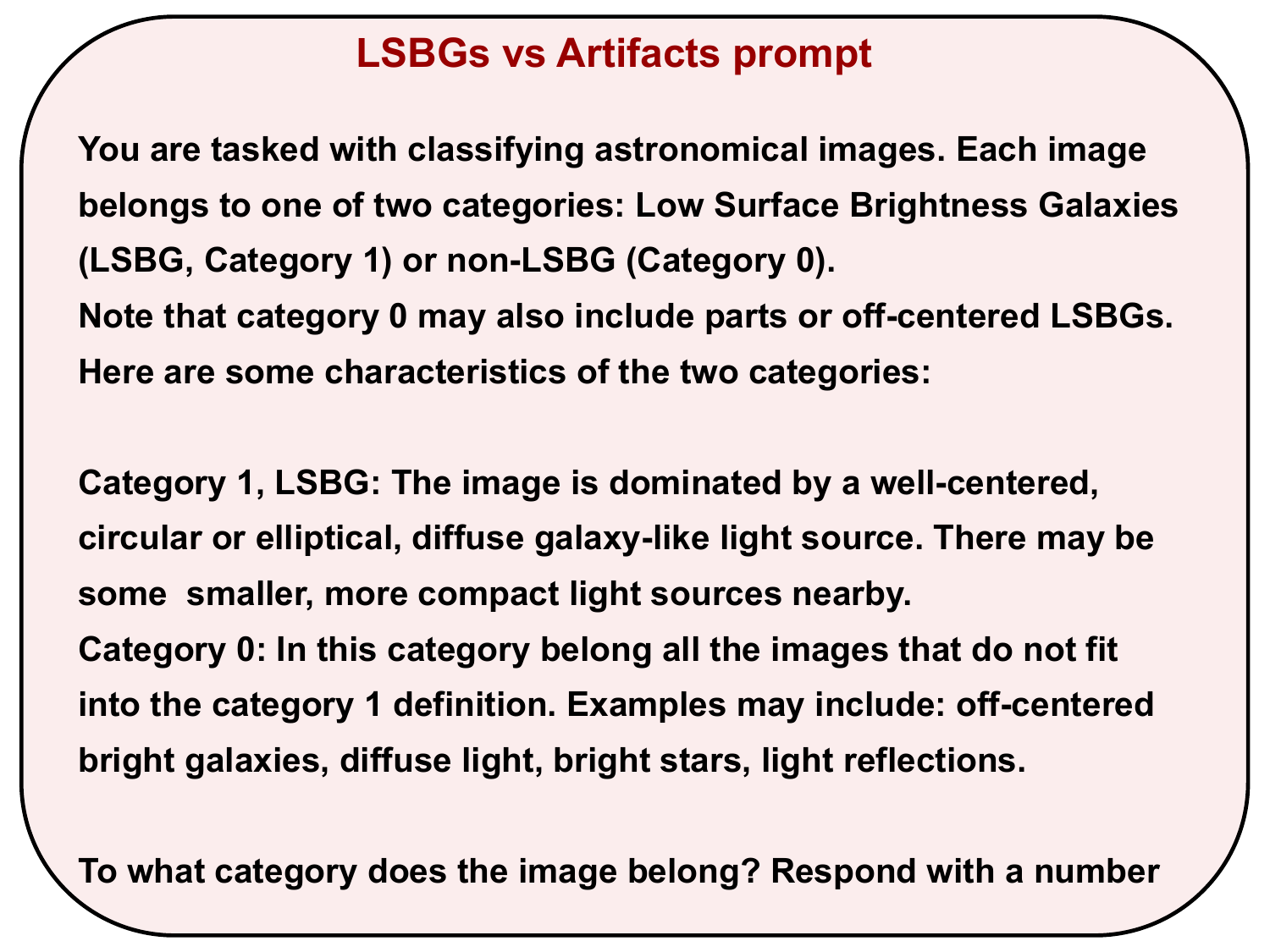}
    \vspace{1cm}\hspace{0.5cm}
    \includegraphics[width=6.8cm]
    {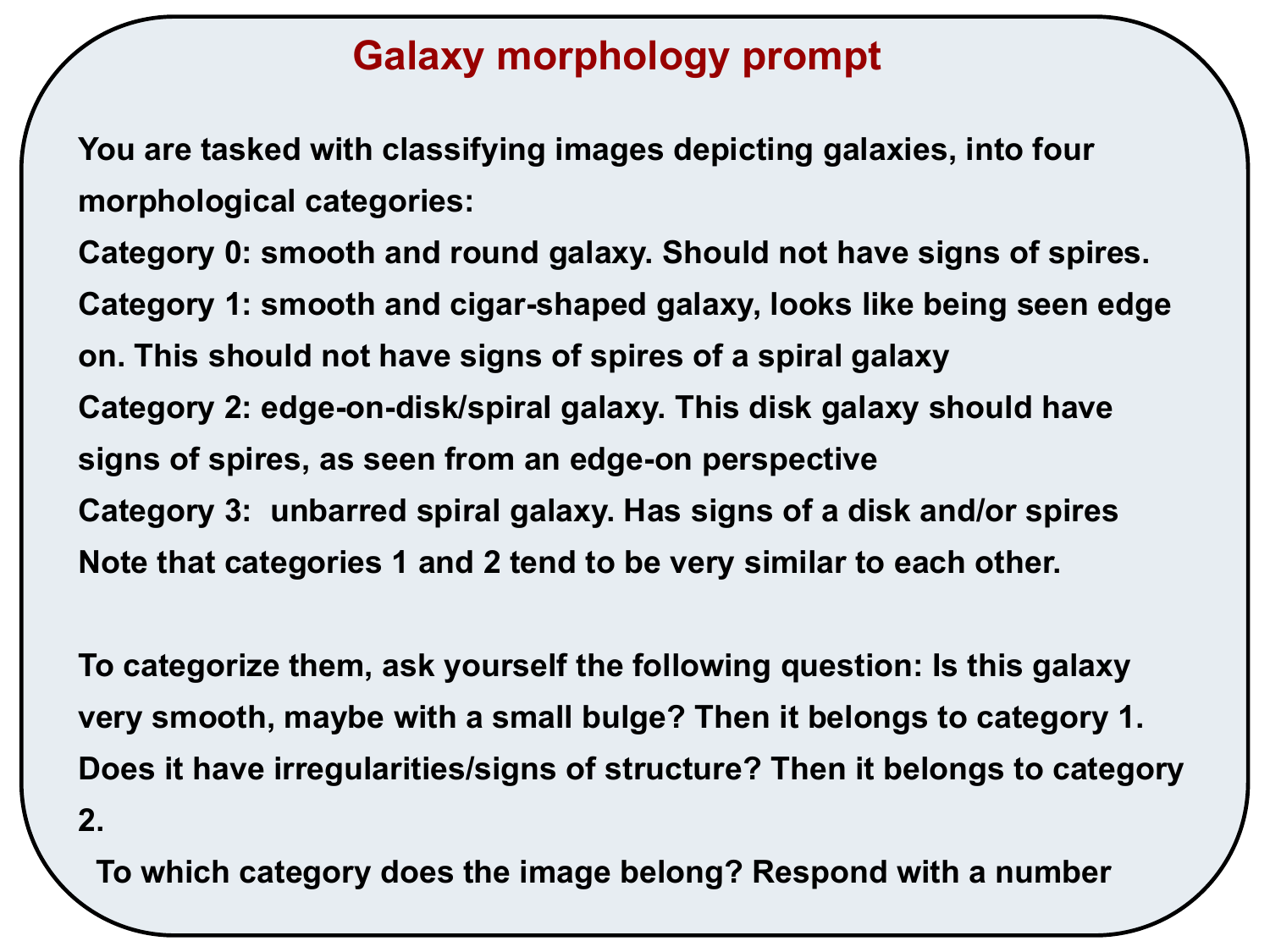}
    \includegraphics[width=6.5cm]{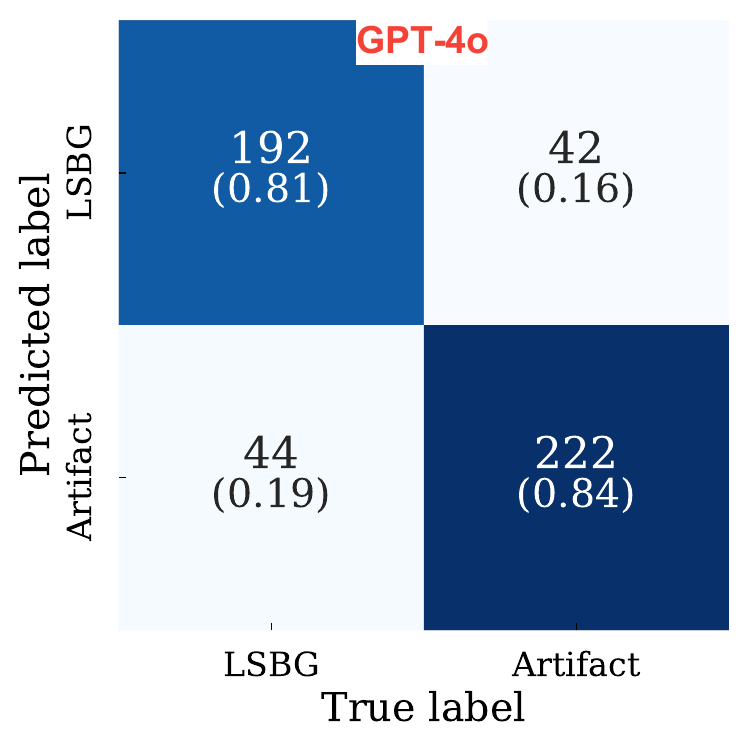}
    \vspace{0.8cm}\hspace{1.0cm}
    \includegraphics[width=6.5cm]{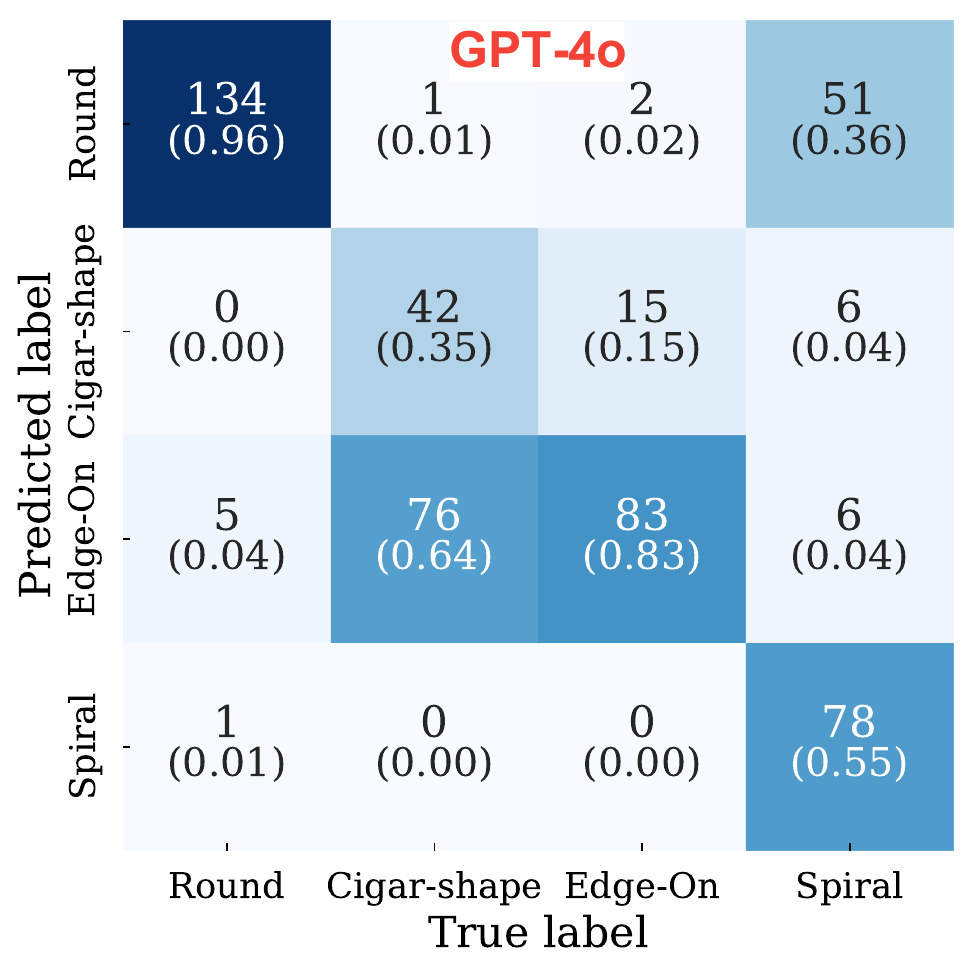}
    \includegraphics[width=6.5cm]{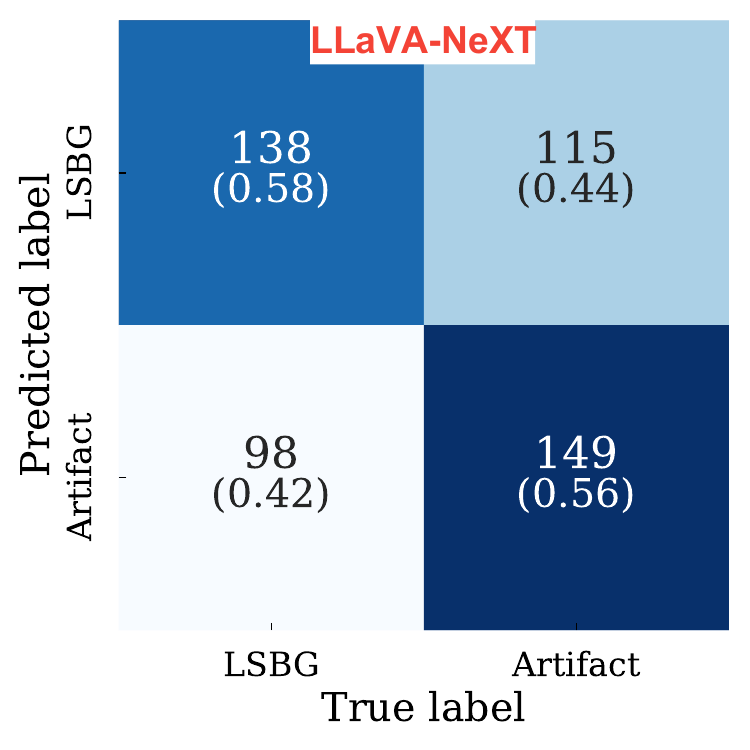}
    \vspace{0.2cm}\hspace{1.0cm}
    \includegraphics[width=6.5cm]{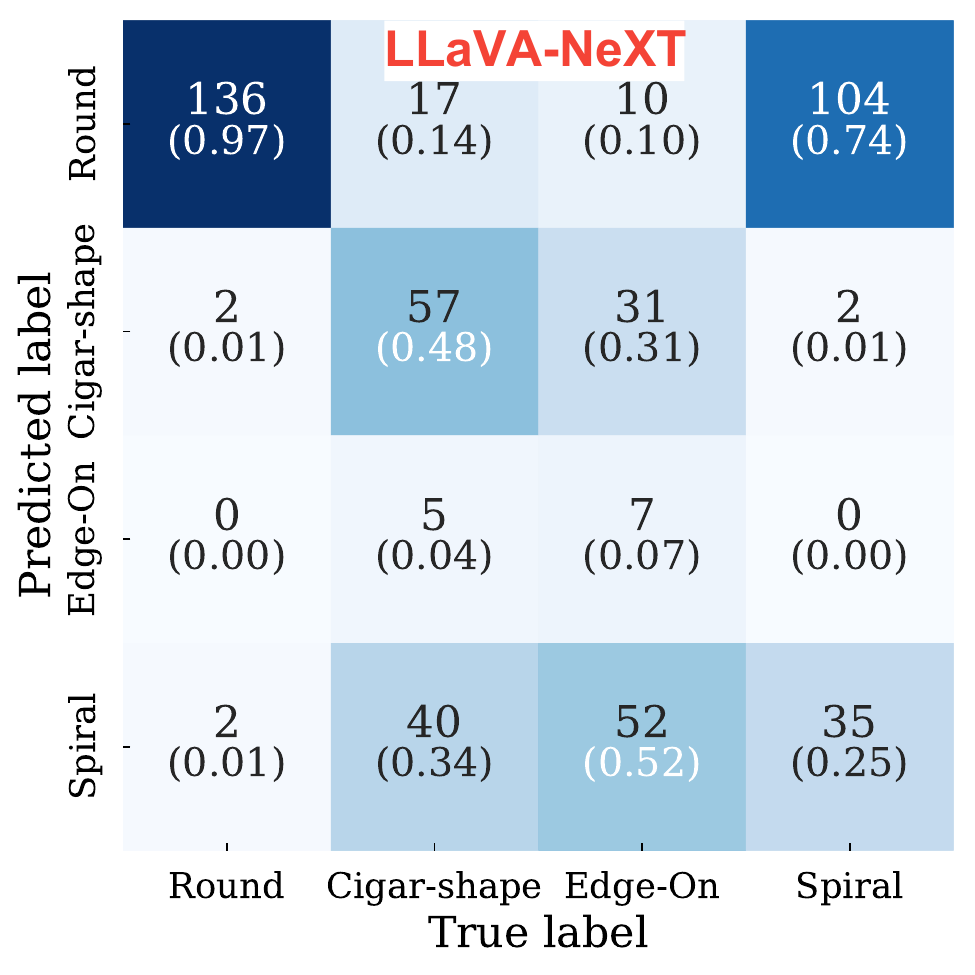}
    \caption{The prompts used to instruct the multimodal models on the two classification tasks (upper part), as well as the resulting confusion matrices, obtained using the GPT-4o (middle part) and LLaVa-NeXT (lower part) models. The confusion matrices compare the predicted vs. the true labels for the example images in our test datasets. The numbers in parentheses correspond to the fraction of images with true label in class $x$ that were predicted to belong to class $y$.}
    \label{fig:Figure}
\end{figure}

We select two problems as examples for our study: the classification of 
Low-Surface-Brightness galaxies (LSBGs) vs artifacts, as described in \citet{Tanoglidis_2020}, and the morphological classification of galaxies into four categories (smooth and round, smooth and cigar-shaped, edge-on disk, and unbarred spiral) using the \texttt{GalaxyMNIST}\footnote{\url{https://github.com/mwalmsley/galaxy_mnist}} dataset  \citep{Decals_2021}.

We use the state-of-the-art, multimodal, GPT-4o model \citep{GPT-4report}, accessed through the OpenAI API. For comparison, we also employ the open-source, language-vision, LLaVA-Next model \citep{LLava-Next}, accessed through the \texttt{Transformers} \citep{Transformers_Library} library from Hugging Face. Specifically, we use the version with Mistral-7B \citep{Mistral_7b} as the LLM backbone, and we perform 4-bit quantization to reduce the memory requirements when loading the model. Both models can process both textual and image inputs (prompts) by creating embeddings for the different modalities in a common embedding space (for this technique see, for example, \citet{CLIP}, and \citet{Sharma_2024} for a recent application in astrophysics).

We want to investigate how natural language descriptions of the different classes can instruct an VLM to accurately classify astronomical images, without further training. For this reason, we provide the exact prompts used in our analysis -- see the upper part of Fig. \ref{fig:Figure}. For each one of the two classification problems, we select a sample of 500 images; we run each model in a loop and we ask it to output only a single number that corresponds to the class the image belongs. The code we used for our experiments, can be found at:
\url{https://github.com/dtanoglidis/zero-shot-astro}.

\section{RESULTS AND DISCUSSION} \label{sec:results}

The results of the zero-shot classification experiments for the two problems described in the previous section are presented in Fig. \ref{fig:Figure}. In the middle (lower) panels we show the confusion matrices (that contrast the predicted vs the actual labels for the sample of images we classified)  from the GPT-4o (LlaVA-NeXT) model. Classifying LSBGs vs artifacts using GPT-4o reaches an accuracy of $\sim 83\%$; this can be compared to the $\sim 92\%$ presented in \citep{Tanoglidis_2020} using custom-trained convolutional neural networks (CNNs) on a training set of 40k examples. Despite the lower performance, we emphasize that this was achieved using the prompt descriptions \textit{alone}, without any further re-training (or the associated coding and choice of CNN architecture). The overall accuracy in the morphological classification problem is lower ($\sim 67\%$), but this is driven mainly by the confusion between cigar-shaped and edge-on galaxies; visual inspection of some examples in Fig. \ref{fig:morphology_examples} shows that it is challenging to distinguish between these two categories. If these two categories are combined into a single one, the overall accuracy rises to $\sim 86\%$.

It is important to note that the classification is not a result of \textit{memorization} (for this problem, see e.g., \citet{memorization}) of data present in the training set: we checked that without providing descriptions of LSBGs and artifacts, GPT-4o tended to classify every image as belonging to a galaxy.

Despite the significantly lower performance of LlaVA-NeXT on both problems (see comparisons between the LLaVA and GPT-4 families at \url{https://llava-vl.github.io/}), we chose to present the results from LLaVA-NeXT since it is an open-source model and can be fine-tuned (e.g. \citet{peft}) in the future for classification tasks on astronomical images and text (see for example the AstroLLaVA model, currently accessible as a Hugging Face space at: \url{https://huggingface.co/spaces/universeTBD/astroLLaVA}). Another technique that may improve the performance of VLMs is the inclusion of image examples from each category in the prompt (few-shot learning). 

This research note aims to demonstrate to the astronomy community the power of multimodal models, particularly Vision-Language Models (VLMs), in addressing complex classification tasks. While our current results are not yet competitive with those achieved by supervised models, they highlight the significant potential of VLMs for various astronomical applications (for a discussion of different applications and tasks that LVMs can perform, see, for example \citet{Dawn_LMMs}). With ongoing advancements in generative AI, fine-tuning or developing custom-built models with increased astronomical domain knowledge, we anticipate that LVMs will play a crucial role in automating and improving astronomical data analysis in the near future.

Furthermore, applications for pedagogical exercises are promising as well. At both the high school and undergraduate level, analysis of images from Hubble, Webb, Subaru and other telescopes can make for appealing exercises. A multimodal model can provide a powerful first layer to give students easy-to-use images from full field data, so students can then carry out scientific analyses such as studies of spiral arms or bars in disk galaxies.

%However, currently such exercises require ``preparing'' the data. With the example prompts shown above, it is easy to imagine exercises where students choose images to analyse from web archives and obtain cut-out images of desired objects -- galaxies with a certain morphology, bright stars, nebulae and so on. Once GPT-4o or equivalent has created a set of images, students can then carry out scientific analyses such as studies of spiral arms or bars in disk galaxies. Thus multi-modal models can provide a powerful first layer to provide students with easy to use images from full field data. Numerous other applications are possible. 

%% IMPORTANT! The old "\acknowledgment" command has be depreciated. It was
%% not robust enough to handle our new dual anonymous review requirements and
%% thus been replaced with the acknowledgment environment. If you try to 
%% compile with \acknowledgment you will get an error print to the screen
%% and in the compiled pdf.
%% 
%% Also note that the akcnowlodgment environment does not support long amounts of text. If you have a lot of people and institutions to acknowledge, do not use this command. Instead, create a new \section{Acknowledgments}.
\begin{acknowledgments}
We thank Marc Huertas-Company, Josh Nguyen and Helen Qu for helpful discussions, Lily Noyes for feedback on explorations of VLMs, and the UniverseTBD team (\url{https://universetbd.org/}) for information about their upcoming AstroLLaVA model. 
\end{acknowledgments}

\vspace{5mm}

%% Similar to \facility{}, there is the optional \software command to allow 
%% authors a place to specify which programs were used during the creation of 
%% the manuscript. Authors should list each code and include either a
%% citation or url to the code inside ()s when available.

\software{Open GPT-4o API \citep{GPT-4report}, \texttt{Transformers} \citep{Transformers_Library},  \texttt{bitsandbytes} \citep{bits_bytes}
          }

%% Appendix material should be preceded with a single \appendix command.
%% There should be a \section command for each appendix. Mark appendix
%% subsections with the same markup you use in the main body of the paper.

%% Each Appendix (indicated with \section) will be lettered A, B, C, etc.
%% The equation counter will reset when it encounters the \appendix
%% command and will number appendix equations (A1), (A2), etc. The
%% Figure and Table counter will not reset.

\appendix
\label{sec:appendix}

In this appendix we present example images from the different classes considered in the classification problems mentioned in the main text. 

\begin{figure}[ht!]
\plottwo{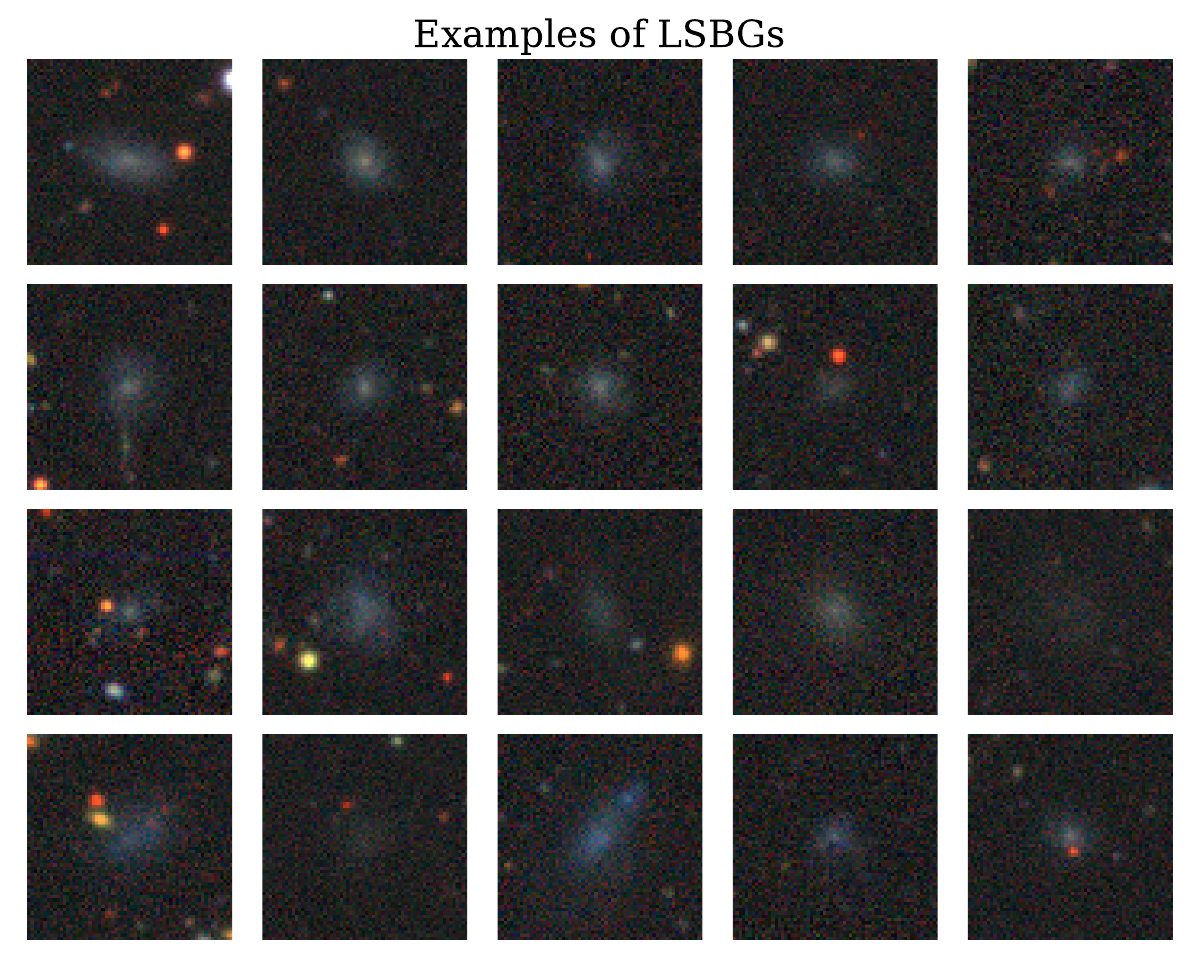}{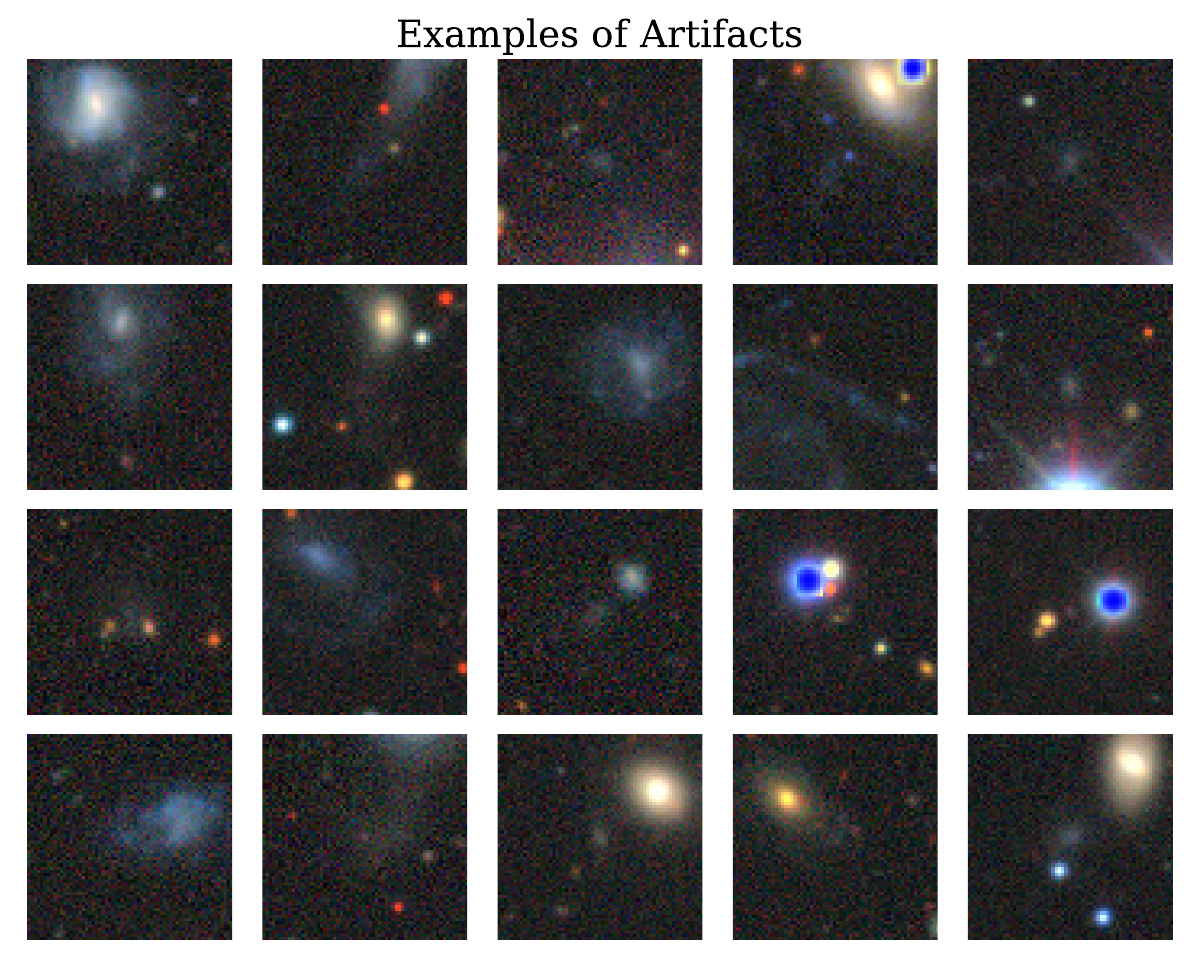}
\caption{Examples of LSBG and artifact images.
\label{fig:LSBG_examples}}
\end{figure}

\begin{figure}[ht!]
\centering
\includegraphics[width=9cm]{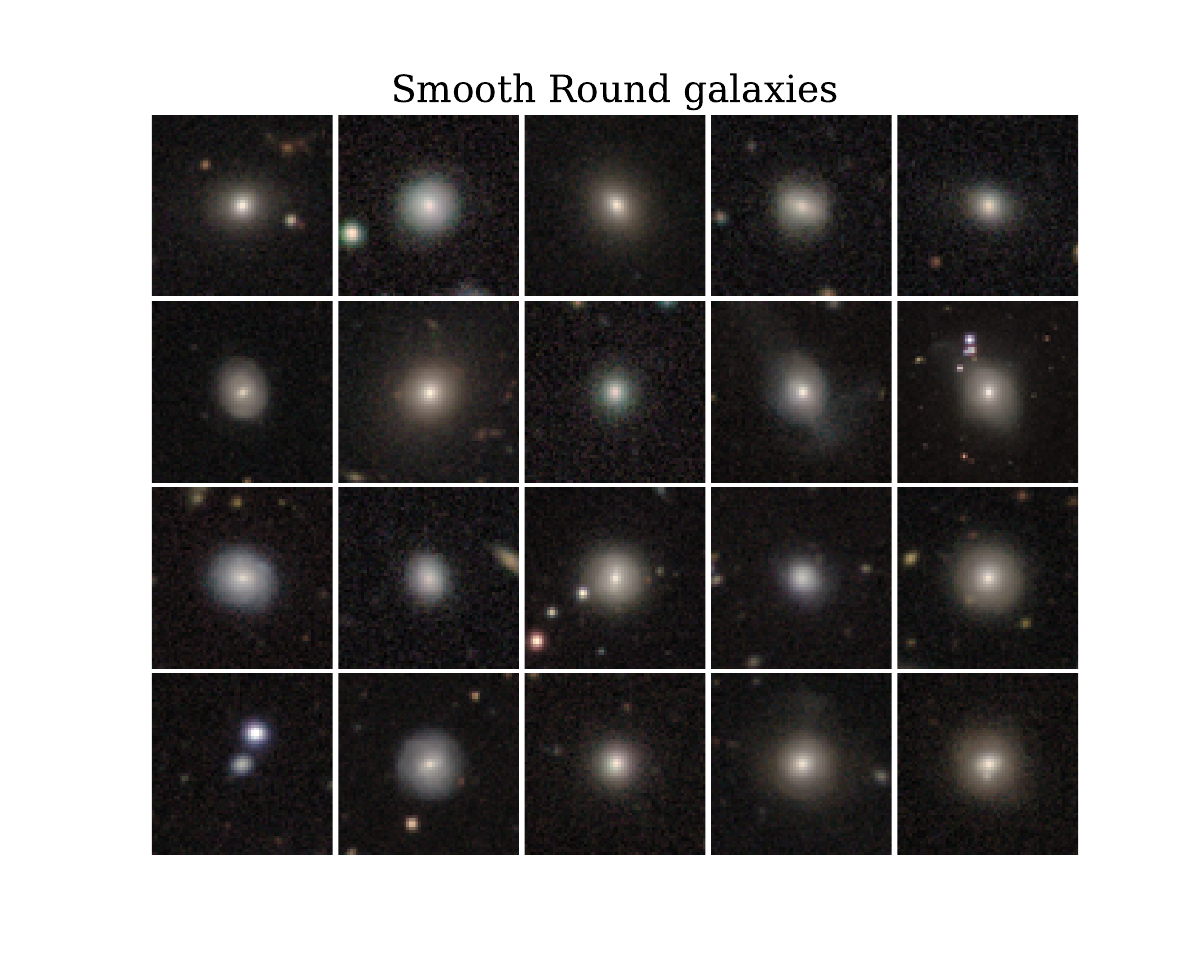}\hspace{-1cm}
\includegraphics[width=9cm]{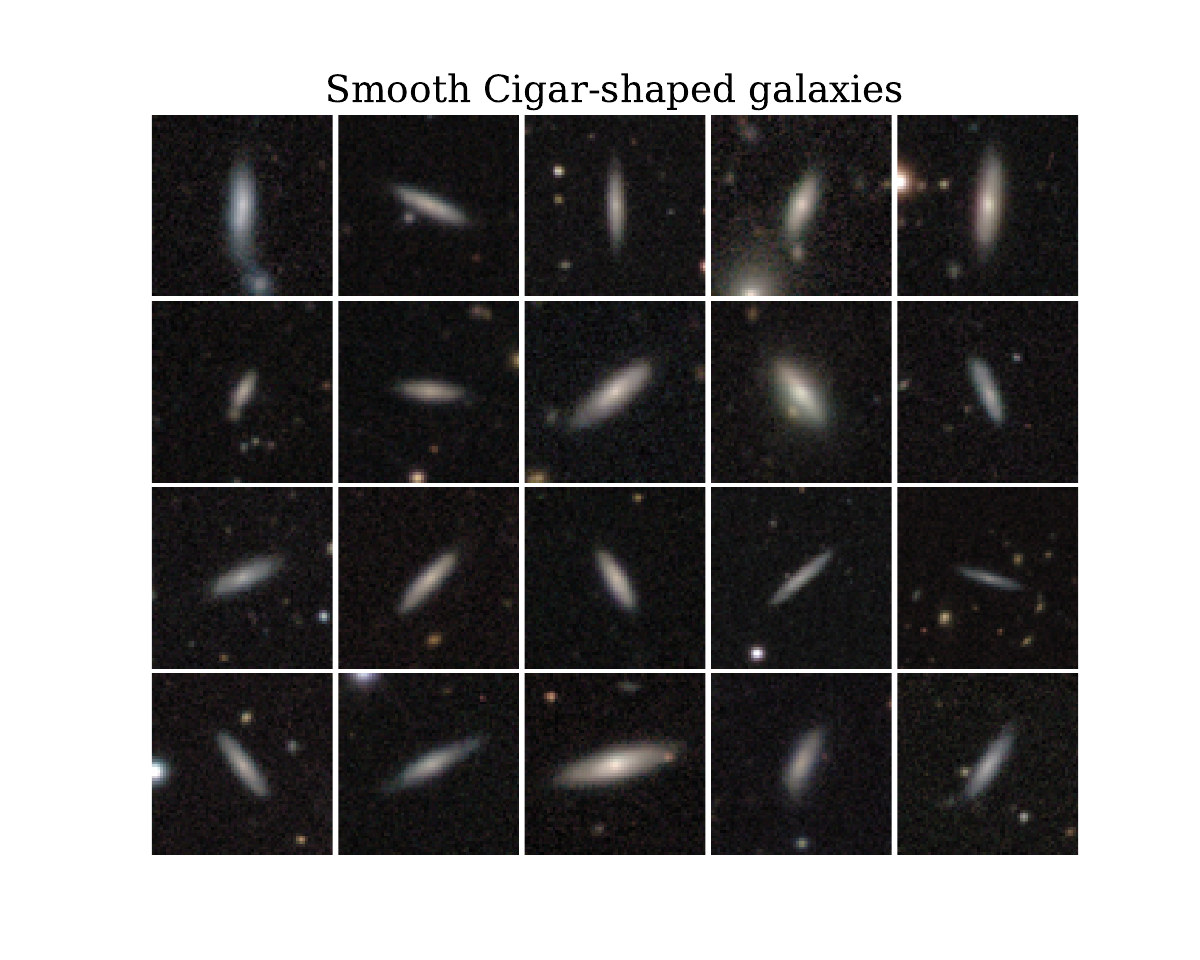}
\includegraphics[width=9cm]{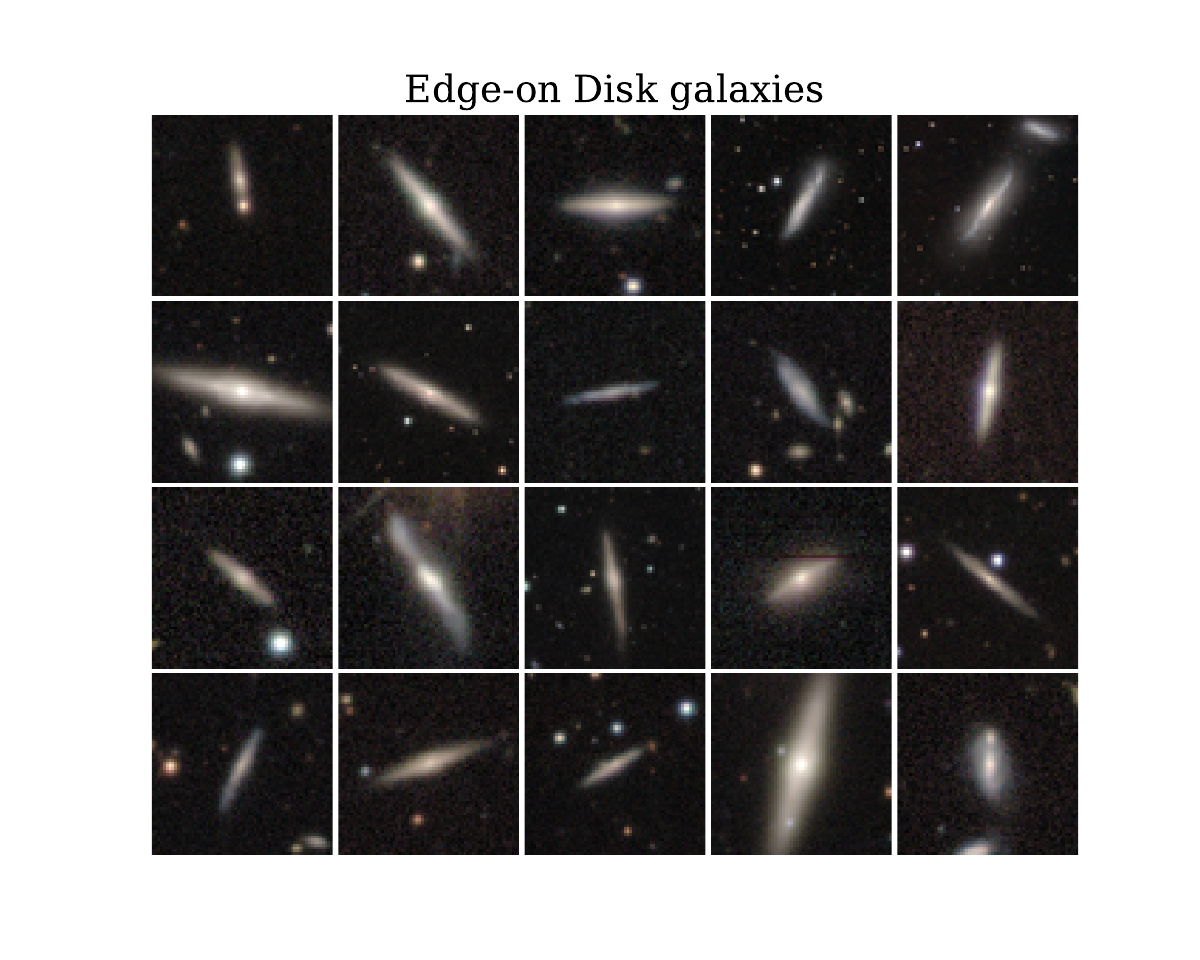}\hspace{-1cm}
\includegraphics[width=9cm]{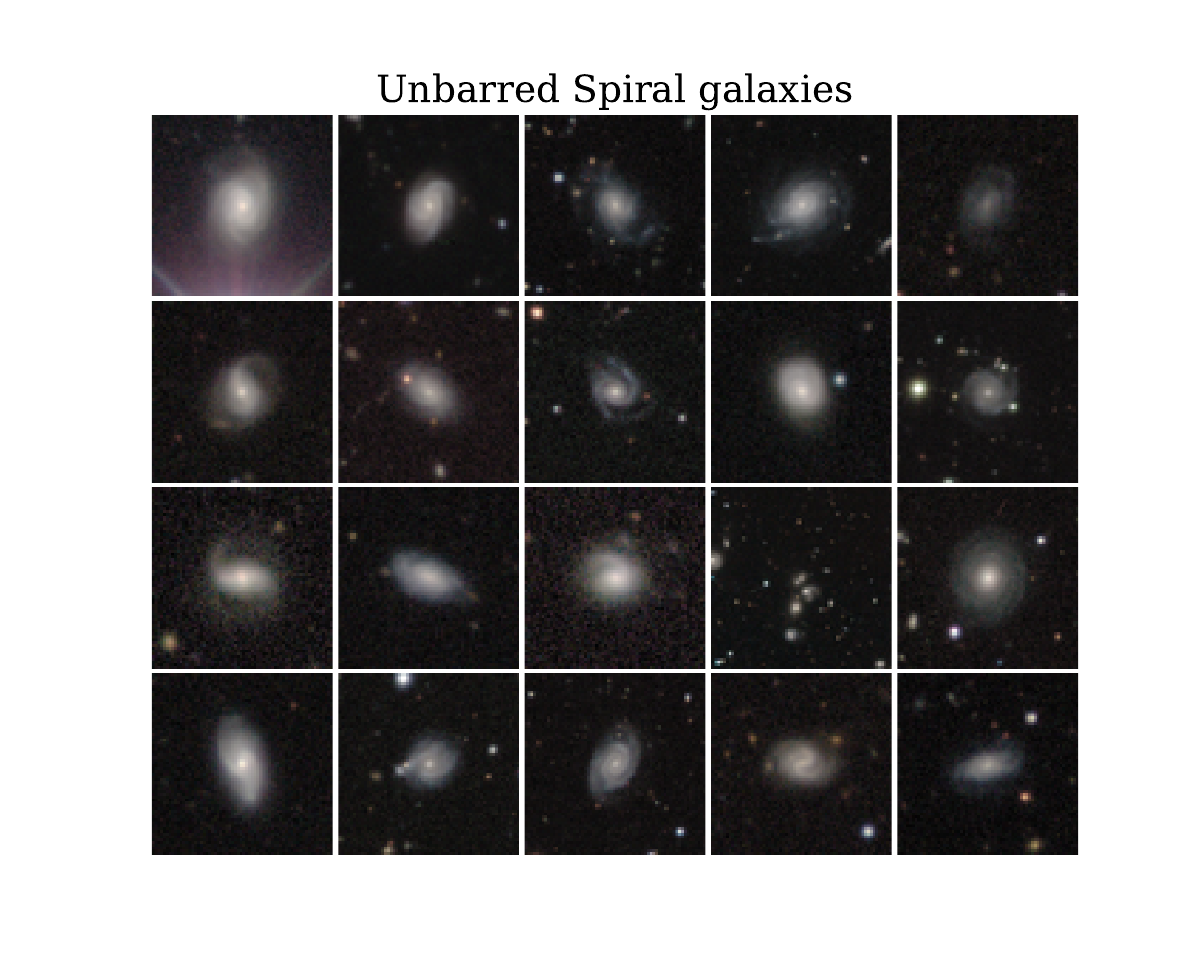}
\caption{Examples of images from the four morphological categories in the GalaxyMNIST dataset. Notice the visual similarity between the smooth cigar-shaped and the edge-on disk galaxies. 
\label{fig:morphology_examples}}
\end{figure}

%% For this sample we use BibTeX plus aasjournals.bst to generate the
%% the bibliography. The sample631.bib file was populated from ADS. To
%% get the citations to show in the compiled file do the following:
%%
%% pdflatex sample631.tex
%% bibtext sample631
%% pdflatex sample631.tex
%% pdflatex sample631.tex

\bibliography{sample631}{}
\bibliographystyle{aasjournal}

%% This command is needed to show the entire author+affiliation list when
%% the collaboration and author truncation commands are used.  It has to
%% go at the end of the manuscript.
%\allauthors

%% Include this line if you are using the \added, \replaced, \deleted
%% commands to see a summary list of all changes at the end of the article.
%\listofchanges

\end{document}